\documentclass[twocolumn]{article}
\usepackage{epsf,epsfig,ifthen,array,latexsym,url}
\usepackage{amsmath,amstext,amsfonts,amssymb}
\usepackage{psfrag}

\makeatletter
\def\@normalsize{\@setsize\normalsize{12pt}\xpt\@xpt
\abovedisplayskip 10pt plus2pt minus5pt\belowdisplayskip \abovedisplayskip
\abovedisplayshortskip \z@ plus3pt\belowdisplayshortskip 6pt plus3pt
minus3pt\let\@listi\@listI} 

\def\subsize{\@setsize\subsize{12pt}\xipt\@xipt}

\def\section{\@startsection {section}{1}{\z@}{24pt plus 2pt minus 2pt}
{12pt plus 2pt minus 2pt}{\large\bf}}

\def\subsection{\@startsection {subsection}{2}{\z@}{12pt plus 2pt minus 2pt}
{12pt plus 2pt minus 2pt}{\subsize\bf}}
\makeatother

\date{}

\psfrag{s0}{{\scriptsize$s_0$}}
\psfrag{s1}{{\scriptsize$s_1$}}
\psfrag{s2}{{\scriptsize$s_2$}}
\psfrag{s3}{{\scriptsize$s_3$}}
\psfrag{s4}{{\scriptsize$s_4$}}
\psfrag{s5}{{\scriptsize$s_5$}}
\psfrag{s6}{{\scriptsize$s_6$}}
\psfrag{s7}{{\scriptsize$s_7$}}
\psfrag{s8}{{\scriptsize$s_8$}}
\psfrag{oB}{{$\o B$}}

\title{Events in Linear-Time Properties}

\author{
  \begin{tabular}[t]{c@{\extracolsep{2em}}c}
  Dimitrie O. P\u{a}un & Marsha Chechik
  \end{tabular}\\
  Department of Computer Science\\
  University of Toronto\\
  Toronto, ON M5S 3G4, Canada\\
  \texttt{\{dimi,chechik\}@cs.toronto.edu}
}
\date{\empty}

\begin{document}
\font\bigmath=cmsy10 scaled 3000
\newtheorem{definition}{Definition}
\newtheorem{theorem}{Theorem}
\newtheorem{prop}{Property}

\newcommand{\reminder}[1]
  {\noindent$|$ \marginpar{\mbox{$|\Leftarrow$ TODO}}#1}
\newcommand{\NM}{\mathit}
\newcommand{\nm}{\mathit}

\newcommand{\limp}{\Rightarrow}
\newcommand{\lpmi}{\Leftarrow}
\renewcommand{\leq}{=}
\newcommand{\lif}[3]{\mathord{\mathbf{if}\ #1\ \mathbf{then}\ #2\ \mathbf{else}\ #3}}

\newcommand{\largeboldbinop}[1]{\mathbin{\mbox{\large$\boldsymbol{#1}$}}}
\newcommand{\Limp}{\largeboldbinop{\limp}}
\newcommand{\Lpmi}{\largeboldbinop{\lpmi}} 
\newcommand{\Leq}{\largeboldbinop{\leq}}

\newcommand{\quant}[3]{\mathord{#1} #2 \in #3 \cdot} 
\newcommand{\E}{\quant{\exists}}   
\newcommand{\A}{\quant{\forall}}   

\renewcommand{\a}{\mathord{\Box}}               
\newcommand{\e}{\mathord{\Diamond}}             
\newcommand{\U}{\mathbin{\mathcal{U}}}          
\newcommand{\W}{\mathbin{\mathcal{W}}}          
\renewcommand{\o}{\mathord{\bigmath\circ}}      
\newcommand{\eu}{\mathord{\uparrow}}            
\newcommand{\ed}{\mathord{\downarrow}}          
\newcommand{\ea}{\mathord{\updownarrow}}        
\newcommand{\cusleft}{\mathopen{\ll}}
\newcommand{\cusright}{\mathclose{\gg}}
\newcommand{\cus}[1]{\cusleft #1 \cusright}

\newcommand{\llbrack}{\mathopen{\lbrack\!\lbrack}}
\newcommand{\rrbrack}{\mathclose{\rbrack\!\rbrack}}
\newcommand{\interp}[2]{#2 \llbrack #1 \rrbrack}

\newcommand{\removestutter}[1]{\natural #1}
\newcommand{\rep}[2][]{\ifthenelse{\equal{#1}{}}{* #2}{#1\!*\!#2}}


\newcommand{\states}{\mathbf{St}}          
\newcommand{\stseq}{\mathbf{St}^\infty}    
\newcommand{\nat}{\mathbb{N}}

\newcommand{\nil}{\NM{nil}}
\newcommand{\FALSE}{\bot}
\newcommand{\TRUE}{\top}

\newenvironment{LTL}  %
{\begin{displaymath}\begin{array}{l}} %
{\end{array}\end{displaymath}}

\newenvironment{proof} %
{\[\begin{array}{cl>{$}l<{$}}} %
{  \end{array}\]}
\newcommand{\lhint}[1]{\\ &\rule{0cm}{3ex}\text{#1} \\ }

\maketitle

\thispagestyle{empty}

\begin{center}
{\large\bf Abstract}
\end{center}
{\em For over a decade, researchers in formal methods tried to create
formalisms that permit natural specification of systems and
allow mathematical reasoning about their correctness.  The
availability of fully-automated reasoning tools enables more
non-specialists to use formal methods effectively --- their
responsibility reduces to just specifying the model and expressing the
desired properties.  Thus, it is essential that these properties be
represented in a language that is easy to use and sufficiently
expressive.

Linear-time temporal logic~\cite{manna89} is a formalism that has been
extensively used by researchers for specifying properties of systems.
When such properties are {\em closed under stuttering}, i.e. their
interpretation is not modified by transitions that leave the system in
the same state, verification tools can utilize a partial-order
reduction technique~\cite{holzmann97} to reduce the size of the model
and thus analyze larger systems.  If LTL formulas do not contain the
``next'' operator, the formulas are closed under stuttering, but the
resulting language is not expressive enough to capture many important
properties, e.g., properties involving events.  Determining if an
arbitrary LTL formula is closed under stuttering is hard --- it has
been proven to be PSPACE-complete~\cite{peled96}.

In this paper we relax the restriction on LTL that guarantees closure
under stuttering, introduce the notion of \emph{edges} in the context
of LTL, and provide theorems that enable syntactic reasoning about
closure under stuttering of LTL formulas.}

\section{Introduction}
\label{intro}

Formal specification of systems has been an active area of research
for several decades.  From finite-state machines to process algebras
to logics, researchers try to create formalisms that would permit
natural specification of systems and allow mathematical reasoning
about their correctness.  However, most of these formalisms have not
been adopted widely outside academia --- their cost-saving benefits
were doubtful, they lacked tool support, and were perceived difficult
to apply~\cite{rosenblum96}.

Recently, the tools for proving properties of finite-state models are
becoming increasingly available and are often used for analyzing
requirements,
e.g.~\cite{atlee93a,bharadwaj99,easterbrook98a,chechik98a}.  These
tools typically require the users to specify properties using temporal
logics and to describe models of systems using some finite-state
transition representation.  The tools are based on a variety of
verification techniques.  For example, SPIN~\cite{holzmann97} and
SMV~\cite{mcmillan93} are based on state-space exploration, also
called \emph{model-checking}, Concurrency
Workbench~\cite{cleaveland93} on bisimulation, and
COSPAN~\cite{harel90} on language containment.  Most finite-state
verification techniques can be fully automated, and the responsibility
of the user reduces to just specifying the model and expressing the
desired properties.  In this context, it is important that properties
can be represented in a language that is easy to use and sufficiently
expressive, to enable even fairly novice users to use it effectively.

Linear-time logic (LTL)~\cite{manna89} is a temporal logic that has
been extensively used by researchers for specifying properties of
systems.  A highly desirable property of LTL formulas is that they are
\emph{closed under stuttering}~\cite{abadi91}.
In particular, the mechanical analysis of such formulas, such as by
the model-checker SPIN~\cite{holzmann97}, can utilize powerful
partial-order reduction algorithms that can dramatically reduce the
state-space of the model. Unfortunately, closure under stuttering can
be guaranteed only for a subset of LTL~\cite{lamport94}, and this
subset is not expressive enough to represent even fairly simple
properties, such as:
\begin{quote}
  \emph{The magnet of the crane may be deactivated only when the
  magnet is above the feed belt.}
\end{quote}
Users of LTL typically try to remedy this problem by introducing extra
variables inside the model --- a technique which tends to clutter the
model, enlarge the state space, and introduce errors.

Determining whether an LTL formula is closed under stuttering is hard:
the problem has been shown to be PSPACE-complete~\cite{peled96}.  Even
though a complete solution is impractical, we have been able to
categorize a subset that is \emph{useful in practice}.  In particular,
a computationally feasible algorithm which can identify a subclass of
closed under stuttering formulas has been proposed
in~\cite{holzmann96} but not yet implemented in SPIN.  The algorithm
is fairly sophisticated and cannot be applied by hand.  Moreover, it
is not clear how often the subclass of formulas identified by the
algorithm is encountered in practice.

In this paper we relax the restriction on LTL that guarantees closure
under stuttering, introduce the notion of \emph{edges} in the context
of LTL, and provide theorems that enable syntactic reasoning about
closure under stuttering of LTL formulas.  
The rest of the paper is organized
as follows: Section~\ref{bg} provides some background on LTL and the
notation used throughout the paper.  Section~\ref{cusedges} discusses
closure under stuttering and introduces edges.  Section~\ref{props}
gives some important properties of edges and closure under
stuttering. Section~\ref{patterns} describes an application of this
work to property patterns identified by Dwyer and his colleagues
in~\cite{dwyer98a}. We discuss some alternative approaches in
Section~\ref{discussion} and conclude the paper in
Section~\ref{conclusion}.

\section{Background}
\label{bg}

We begin by briefly introducing our notation which we have adopted
from~\cite{hehner93}. A sequence (or string) is a succession of
elements joined by semicolons.  For example, we write the sequence
composed of the first five natural numbers, in order, as $0;1;2;3;4\ $
or, more compactly, as $0;..5$ (note the left-closed, right-open
interval).  We can obtain an item of the sequence by subscripting:
$(0;2;4;5)_2 = 4$.  When the subscript is a sequence, we obtain a
subsequence: $(0;3;1;5;8;2)_{1;2;3} = (0;3;1;5;8;2)_{1;..4} = 3;1;5$.

A state is modeled by a function that maps variables to their values,
so the value of variable $a$ in state $s_0$ is $s_0(a)$. We denote the
set of all infinite sequences of states as $\stseq$, and the set of
natural numbers as $\nat$.

Boolean expressions are connected by $\lnot$(not), $\land$ (and),
$\lor$ (or), $\limp$ (implies), $\lpmi$ (is implied by), and $=$ (if
and only if). To reduce the clutter of parenthesis, we denote the main
connective in a formula by a bigger and bolder symbol, e.g. $\Leq$. We
consider the connectives to have decreasing precedence as follows:
$=$, $\lnot$, $\land$, $\lor$, $\limp$; the connectives $\Leq$,
$\Limp$, and $\Lpmi$ have the lowest precedence. For example, the
formula \( a \land b \lor b \Leq a \limp b \) should be parsed as:
\(((a\land b)\lor b)\Leq(a \limp b)\).

Linear time temporal logic (LTL) is a language for describing and
reasoning about sequences of states. These sequences can be
interpreted in a variety of ways: the state of the world as it evolves
over time, the state of a program as it is executing, and so
forth. Informally, LTL is comprised of propositional formulas and
temporal connectives $\a$ (always), $\e$ (eventually), $\o$ (next),
and $\U$ (until). The first three operators are unary, while the last
one is binary. Using these operators, one can express properties about
the evolution of a system. For example, a formula $p \lor \o q$
indicates that either $p$ holds in the starting state of the system,
or $q$ holds in the following state.  A formula $\a(p \limp \e q)$
indicates that each occurrence of $p$ is followed, at some point, by
an occurrence of $q$. We now define the formal semantics of an LTL
formula, based on its syntactic structure. Let $A$ and $B$ be LTL
formulas, let $a$ be a variable name, and let $s$ be a sequence of
states, i.e.  $s\in\stseq$. We denote an interpretation of formula
$F$ in state sequence $s$ as $\interp{F}{s}$, and define it as
follows:
\begin{eqnarray*}
  \interp{a}{s}       &\Leq& s_0(a) 
  \label{eq:Svar} \\
  \interp{\lnot A}{s}  &\Leq& \lnot \interp{A}{s}
  \label{eq:Snot} \\
  \interp{A \land B}{s}&\Leq& \interp{A}{s} \land \interp{B}{s}
  \label{eq:Sand} \\
  \interp{\o A}{s}    &\Leq& \interp{A}{s_{1;..\infty}}
  \label{eq:So} \\
  \interp{\a A}{s}    &\Leq& \A{i}{\nat}\interp{A}{s_{i;..\infty}} 
  \label{eq:Sa} \\
  \interp{\e A}{s}    &\Leq& \E{i}{\nat}\interp{A}{s_{i;..\infty}}
  \label{eq:Se} \\
  \interp{A \U B}{s}  &\Leq& \E{i}{\nat}(\interp{B}{s_{i;..\infty}}\land 
  \nonumber \\
                      &&\hfil\A{j}{\{0,1,..,i-1\}}\interp{A}{s_{j;..\infty}})
  \label{eq:SU}
\end{eqnarray*}
For example, $\interp{\e A}{s}$ means that a formula $A$ must hold at
some point $i$ in the sequence $s$. We do not describe other boolean
operators as they can be expressed in terms of negation and
conjunction.

We say that an LTL formula $F$ is closed under stuttering when its
interpretation remains the same under state sequences that differ only
by repeated states. We denote a closed under stuttering formula as
$\cus{F}$, and formally define it as follows:
\begin{definition} \label{eq:cus} $\cus{F} \Leq$
\begin{displaymath}
  \A{s}{\stseq}\A{i}{\nat}\interp{F}{s} =
                                \interp{F}{(s_{0;..i};s_i;s_{i;..\infty})}
\end{displaymath}
\end{definition}
In other words, given any state sequence $s$, we can repeat any of its
states without changing the interpretation of $F$. Note that
$s_{0;..i};s_i;s_{i;..\infty}$ is a sequence of states that differ
from $s$ only by the repeated state $s_i$.

It is easy to see that any LTL formula that does not contain the
``$\o$'' operator is closed under stuttering. For example, $\a a$ is
closed under stuttering because no matter how much we repeat states,
we cannot change the value of $a$. On the other hand, the formula $\o
a$ is not closed under stuttering. We can see that by considering the
state sequence $s$ in which $s_0(a)$ is true and $s_1(a)$ is
false. Then $\o a$ is false when we evaluate it in $s$, and true when
we evaluate it in $s_0;s$.

\section{Closure Under Stuttering and Edges}
\label{cusedges}

\begin{figure}[ht] 
\begin{center}
  \epsfig{file=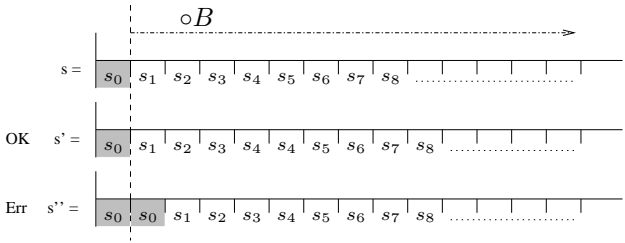,width=\linewidth} 
\end{center}
\caption{Effects of stuttering on formula $\o B$.} \label{fig:oB} 
\end{figure}

Our aim is to capture a large set of formulas that use the ``$\o$''
operator but are still closed under stuttering. We begin by developing
the intuition behind the main theorem that enables us to generate this
class of formulas.  Let $B$ be an LTL formula that is closed under
stuttering.  Consider the formula $\o B$: it may not be closed under
stuttering; furthermore, we can not correct this by simply quantifying
$\o B$ with `$\a$' or `$\e$'. To understand how stuttering can modify
the interpretation of $\o B$, we illustrate two different stuttering
scenarios in Figure~\ref{fig:oB}, where $s$ is a sequence of states,
and $s'$ and $s''$ are derived from $s$ by stuttering the fifth and
the first states respectively. Note that the type of stuttering
exemplified in $s'$, that is, repeating any part of the sequence with
the exception of the first state, causes no harm because $B$ itself is
closed under stuttering.  However, stuttering the first state, as
shown in $s''$, may cause problems, because $\o B$ is evaluated on a
state sequence that includes the new state.

To circumvent this problem, we conjoin $\o B$ with another formula,
$F$, and quantify the resulting expression with `$\e$': $\e(F \land \o
B)$. We need the quantification to ``follow'' the state $s_1$, as
$s_0$ can be stuttered any number of times, while the conjunct $F$ is
needed to ignore the leading sequence of repeated states
$s_0$.\footnote{Note that we could not have used $\a$ for
quantification as it distributes over conjunction. The choice of a
quantifier is dictated by the boolean operator between $F$ and $\o
B$.}  A possible way to determine when a state is repeated is to check
that an LTL formula $A$ that is closed under stuttering does not
change its interpretation from the current to the next
state. Therefore, we can define $F$ as $\lnot A \land \o A$. Note that
$F$ is guaranteed to be false only when the sequence starts with a
repeated state.

We now claim that the formula developed above is closed under
stuttering:
\begin{theorem}  \label{thm:main}
\begin{displaymath}
\cus{A} \land \cus{B} \Limp  \cus{\e(\lnot A \land \o A \land \o B)}  
\end{displaymath}
\end{theorem}
\textbf{Proof Sketch:} Let $s,s' \in \stseq$, such that $s'$ is the
same as $s$ except that the first state of $s'$ is not stuttered. In
other words, we can repeat the first state of $s'$ a number of times
and obtain $s$. We can rewrite the formula $\e(\lnot A \land \o A \land
\o B)$ as  $\e(\lnot A \land \o G)$, where $G = A \land B$. If we let 
$F=\e(\lnot A \land \o G)$, we can see that $\interp{F}{s} =
\interp{F}{s'}$.  So, the interpretation of $F$ is not affected by
stuttering except perhaps when we stutter the first state $s_0$. We
have argued in the discussion leading to the theorem that this is the
case provided that $A$ and $G$ are closed under stuttering, which is
true here.  Thus, we only need to worry about state sequences $s$
which have the first state stuttered.  However, in such cases the
expression $\lnot A \land \o A$ becomes false because $A$ is closed
under stuttering. This means that the entire preamble of stuttered
states that might change the interpretation of the formula is in fact
ignored, and thus the formula remains closed under stuttering. \hfill
$\blacksquare$

This theorem is the main result that allows us to build formulas that
contain the ``$\o$'' operator and are still closed under stuttering.  The
theorem was enabled by the formula $\lnot A \land \o A$,
which expresses a \emph{change} in
$A$. More exactly, it expresses an \emph{up edge} in $A$. We
denote an up edge by $\eu$, a down edge by $\ed$, and an up or down
edge by $\ea$. Their formal definitions follow.
\begin{definition} \label{def:edges}
If $A$ is an LTL formula, then

\vskip 0.1in

\begin{tabular}{rclcl}
  $\eu A$ &$\Leq$& $\lnot A \land\o$ A  & $\;$ --- & up or rising edge    \label{eq:eu} \\
  $\ed A$ &$\Leq$& $A \land \o\lnot A$  & $\;$ --- & down or falling edge \label{eq:ed} \\
  $\ea A$ &$\Leq$& $\eu A \lor \ed  A$  & $\;$ --- & any edge             \label{eq:ea}
  \end{tabular}
\end{definition}

The edges are also called \emph{events}, as they capture the same
notion as the events proposed by the Software Cost Reduction (SCR)
researchers~\cite{heninger80, heitmeyer96b}. The SCR events cannot
explicitly include temporal operators, whereas our formalism enables
reasoning about events in arbitrary LTL formulas.  For example, a
formula $\eu\a A$ has a well-defined interpretation in our
language. We also note here a strong analogy between our (logical)
edges and signal edges. Computer engineers have made good use of edges
in electrical signals --- most circuitry is driven by edges, e.g. the
clock.  Edges are also widely used in other engineering disciplines,
e.g. electrical engineering and telecommunications. It is surprising
that we managed to work around them for so long in the model checking
world!

\section{Properties}
\label{props}

In this section we present a few important properties of edges and
closure under stuttering.  We begin by noting that edges are related
by the following formulas:
\begin{eqnarray}
  \eu\lnot A  &\Leq& \ed A    \label{eq:euNot} \\
  \ed\lnot A  &\Leq& \eu A    \label{eq:edNot} \\
  \ea\lnot A  &\Leq& \ea A    \label{eq:eaNot}
\end{eqnarray}
These formulas allow us to switch between the different types of
edges easily.  The following are some general properties of closure
under stuttering:
\begin{eqnarray}
a \text{ is a variable} &\Limp& \cus{a}          \label{eq:cusVar} \\
\cus{A}                 &\Leq&  \cus{\lnot A}    \label{eq:cusNot} \\
\cus{A} \land \cus{B}   &\Limp& \cus{A \land B}  \label{eq:cusAnd} \\
\cus{A}                 &\Limp& \cus{\a A}       \label{eq:cusA} \\ 
\cus{A}                 &\Limp& \cus{\e A}       \label{eq:cusE} \\
\cus{A} \land \cus{B}   &\Limp& \cus{A \U B}     \label{eq:cusU}
\end{eqnarray}
Note that property~(\ref{eq:cusNot}) is an equality indicating that
when a formula is negated, its closure under stuttering property is
preserved.  Property~(\ref{eq:cusAnd}) indicates that if two formulas
are closed under stuttering, then so is their conjunction.  These two
formulas allow us to conclude that
\[ \cus{A} \land \cus{B} \Limp \cus{A * B} \]
where $*$ is any of $\land$, $\lor$, $\limp$, $\lpmi$, or $=$.  Such
properties enable reasoning about closure under stuttering of a formula
by looking at its components.  Finally, formula~(\ref{eq:cusNot})
together with (\ref{eq:euNot}) and~(\ref{eq:edNot}) allows the
interchangeable use of \(\ed\) and \(\eu\) when analyzing properties
of the form
\[ \cus{A} \Limp f(\eu A) \]
Thus, in the rest of the paper we talk only about the $\eu$-edges in the
context of closure under stuttering.

Below we discuss closure under stuttering properties that contain
edges and the ``$\o$'' operator.  The first property is a corollary of
Theorem~\ref{thm:main}:
\begin{prop} \label{eq:cusEe}
  \begin{multline*}
    \cus{A} \land \cus{B} \land \cus{C} \Limp 
    \cus{\e(\eu A \land \o B \land C)} 
  \end{multline*}
\end{prop}
It has two simplified versions:
\[  \cus{A} \land \cus{B} \Limp \cus{\e(\eu A \land B)} \]
and
\[  \cus{A} \land \cus{B} \Limp \cus{\e(\eu A \land \o B)} \]
that we often encountered in practice.  In both cases an
\emph{existence} property is formalized. The formulas say that the
event $\eu A$ must happen and then $B$ holds.  In these versions, $B$
is evaluated right before or right after the event, respectively.
\begin{prop} \label{eq:cusAe}
  \begin{multline*} 
    \cus{A} \land \cus{B} \land \cus{C} \Limp 
    \cus{\a(\eu A \limp \o B \lor C)} 
  \end{multline*}
\end{prop}
This property is logically equivalent to Property~\ref{eq:cusEe}.
Its two simplified versions are
\[  \cus{A} \land \cus{B} \Limp \cus{\a(\eu A \limp B)} \]
and
\[  \cus{A} \land \cus{B} \Limp \cus{\a(\eu A \limp \o B)} \]
They express a \emph{universality} property: whenever the event $\eu
A$ happens, $B$ will hold.  As in the case of Property~\ref{eq:cusEe},
$B$ is evaluated right before or right after the event, respectively.

Consider the following example:
\begin{quote}
  \emph{Items should only be dropped on the table.}
\end{quote}
This property can be formalized as
\[ \a(\ed \NM{hold} \Limp \NM{pos}=\NM{above\_tbl}), \]
where $\NM{hold}$ is a state variable that is true when we hold an
item, and $\NM{pos}$ is a state variable indicating the position.  Note
that an item is considered dropped if we hold it in one state and
do not hold it in the next.  Formally, we express ``dropped'' as
$\NM{hold} \land \o\lnot \NM{hold}$ or as $\ed \NM{hold}$.

The last property deals with the ``until'' operator:
\begin{prop} \label{eq:cusUe}
  \begin{multline*}
    \cus{A} \land \cus{B} \land \cus{C} \land 
    \cus{D} \land \cus{E} \land \cus{F} \\
    \Limp 
    \cus{(\lnot\eu A \lor \o B \lor C) \U (\eu D \land \o E \land F)} 
  \end{multline*}
\end{prop}
There are many simplified expressions for this property which
are omitted here for brevity.  

For example, a property
\begin{quote}
  \emph{Initially, no items should be dropped on the table before
    the operator pushes and releases the GO button.}
\end{quote}
can be formalized as
\[ \lnot\ed \NM{hold} \U \ed \NM{button}, \]
where $\NM{hold}$ has the same meaning as before, and $\NM{button}$ is
a state variable which is true when the button is pressed and false
otherwise.

\begin{figure}[htb]
  \begin{center}
    \epsfig{file=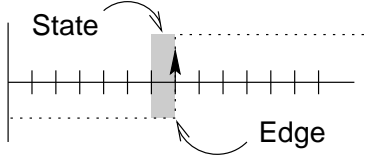} 
    \caption{Edge-detecting state}
    \label{fig:edge}
  \end{center}
\end{figure}
In order to effectively express properties containing edges, it is
important to realize that an edge is detected just \emph{before} it
occurs, as illustrated in Figure~\ref{fig:edge}.\footnote{Note that we
can move the detection of the edge after it occurs if we replace the
``next'' by the ``previous'' operator.}  That is, $\eu A$ becomes true
in the state where $A$ is false.  For example, consider the following
property:
\begin{quote}
  \emph{After the robot deposits an object on the belt, it should not
    hold another object until the sensor at the end of the belt is true.}
\end{quote}
If we formalize it as
\[ \a(\ed\NM{hold} \limp (\lnot\NM{hold} \U \NM{sensor})), \]
we do not get the correct formula since its consequent would be
evaluated one state too early: $\ed\NM{hold}$ is detected when
$\NM{hold}$ is true, but requires $\NM{hold}$ to remain false
until $\NM{sensor}$ is true. The formula can be fixed by considering
the consequent in the ``next'' state:
\[ \a(\ed\NM{hold} \limp \o(\lnot\NM{hold} \U \NM{sensor})) \]

\section{Edges and Patterns}
\label{patterns}

A pattern-based approach to the presentation, codification and reuse
of property specifications for finite-state verification was proposed
by Dwyer and his colleagues in~\cite{dwyer99,dwyer98a}. They
performed a large-scale study in which specifications containing over
500 temporal properties were collected and analyzed. They noticed that
over 90\% of these could be classified under one of the proposed
patterns~\cite{dwyer99}.

We discuss two directions for integrating our work into the pattern
system: extending the system to include events based on edges, and
evaluating the effectiveness of our theorems in determining closure
under stuttering for the newly created, edge-based formulas.  In the
rest of the section we briefly discuss the property pattern system
(following the presentation in~\cite{dwyer98a}), describe our
extensions based on the usage of edges, and conclude with discussing
closure under stuttering.

\subsection{The Pattern System}

The patterns enable non-experts to read and write formal
specifications for realistic systems and facilitate easy conversion of
specifications between formalisms.  Currently, the properties can be
expressed in a variety of formalisms such as LTL, computational tree
logic (CTL)~\cite{clarke86}, quantified regular expressions
(QRE)~\cite{oleander90}, and other state-based and event-based
formalisms.

\begin{figure*}[t]
\begin{center}
\psfig{file=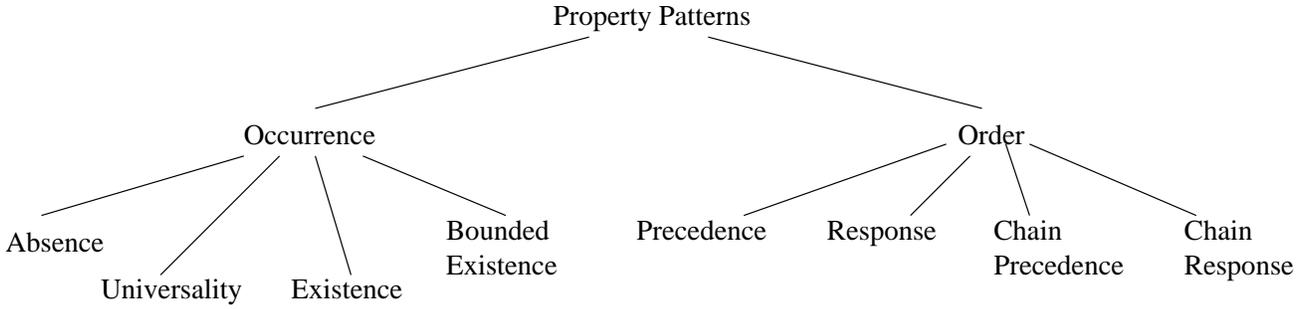,width=\linewidth}
\caption{A Pattern Hierarchy.}
\label{fig:hierarchy}
\end{center}
\end{figure*}

The patterns are organized in a hierarchy based on their semantics, as
illustrated in Figure~\ref{fig:hierarchy}.  Some of the patterns are
described below:
\begin{description}
\item[Absence] A condition does not occur within a scope;
\item[Existence] A condition must occur within a scope;
\item[Universality] A condition occurs throughout a scope;
\item[Response] A condition must always be followed by another
  within a scope;
\item[Precedence] A condition must always be preceded by
  another within a scope.
\end{description}

\begin{figure}[htb]
\begin{center}
\psfig{file=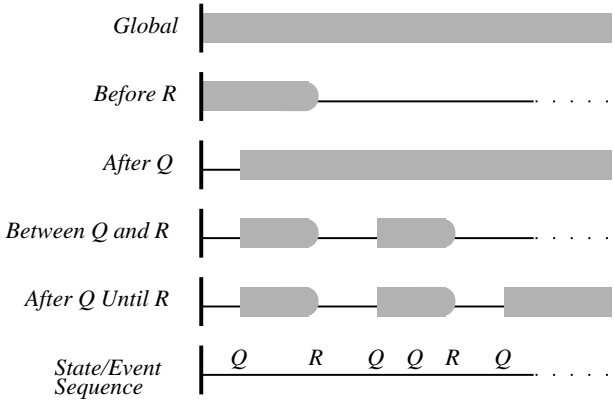,width=\linewidth}
\caption{Pattern Scopes.}
\label{fig:scopes}
\end{center}
\end{figure}

Each pattern is associated with several \emph{scopes} --- the regions
of interest over which the condition is evaluated.  There are five
basic kinds of scopes:
\begin{description}
  \item[A. Global]     The entire state sequence;
  \item[B. Before $R$] The state sequence up to condition $R$;
  \item[C. After $Q$]  The state sequence after condition $Q$;
  \item[D. Between $Q$ and $R$] The part of the state sequence between
    condition $Q$ and condition $R$;
  \item[E. After $Q$ Until $R$] Similar to the previous one except
    that the designated part of the state sequence continues even if the
    second condition does not occur.
\end{description}
These scopes are depicted in Figure~\ref{fig:scopes}. The scopes were
initially defined in~\cite{dwyer99} to be closed-left, open-right
intervals, although it is also possible to define other combinations,
such as open-left, closed-right intervals. 

For example, an LTL formulation of the property ``$S$ precedes $P$
between $Q$ and $R$'' (\textbf{Precedence} pattern with ``between $Q$
and $R$'' scope) is:
\[ \a((Q \land \e R) \limp (\lnot P \U (S \lor R))) \]
Even though the pattern system is
formalism-independent~\cite{dwyer98a}, we are only interested in the
way the patterns are expressed in LTL.

\subsection{Events in LTL Patterns}

It is often natural to express properties using changes of states ---
edges.  As the original pattern system was state-based, we tried to
extend it by incorporating edge-based events. These events can be used
for specifying the conditions as well as for defining the bounding
scopes. For example, we may want to express the properties ``Event $S$
precedes event $P$ between states $Q$ and $R$'', or ``State $S$
precedes state $P$ between events $Q$ and $R$''.  These properties are
often hard to specify correctly, and since all of them contain the
``$\o$'' operator (either directly or indirectly through the use of
edges), it is not trivial to determine if they are closed under
stuttering.

Introducing edges into the patterns generates an explosion in the
number of formulas. The patterns contain up to two conditions, while
the bounding interval has up to two ends; each of these can be either
state-based or edge-based, giving us up to 16 different combinations.
Moreover, when a condition and an interval end are of the same type
(either state-based or edge-based), we can choose to make the interval
open or closed, leading to even more possibilities. Note that when the
condition and the interval end are of different types, there is no
ambiguity because edges occur \emph{between} states.

Due to the large number of possible formulas, we extended only some of
the patterns: \textbf{Absence}, \textbf{Existence},
\textbf{Universality}, \textbf{Precedence}, and \textbf{Response}.  In
the original pattern system, the conditions were represented by
formulas $P$ and $S$, while the bounding interval was defined by
formulas $Q$ and $R$. We have considered the following possibilities:
\begin{flushleft}
  \begin{tabular}{cll}
    0. & $P$, $S$ --- states,  & $Q$, $R$ --- states; \\
    1. & $P$, $S$ --- states,  & $Q$, $R$ --- up edges; \\
    2. & $P$, $S$ --- up edges, & $Q$, $R$ --- states;  \\
    3. & $P$, $S$ --- up edges, & $Q$, $R$ --- up edges.
  \end{tabular}  
\end{flushleft}
Combination 0 corresponds to the original formulation of
\cite{dwyer98a}, where all of $P$, $S$, $Q$ and $R$ are state-based.
The remaining three combinations are our extensions of the pattern
system.  We assume that multiple events can happen simultaneously, but
only consider closed-left, open-right intervals, as in the original
system.  We note, however, that it is perfectly possible to have
formulas for all other combinations of interval bounds.  Down edges
can be substituted for up edges without changing the formulas.

In Figure~\ref{fig:patE} we show the LTL formulations for the
\textbf{Existence} pattern. Each of the scopes is associated
with four formulas corresponding to the four combinations of
state-based and edge-based conditions and interval bounds we have
considered. We have modified several of the 0-formulas
(i.e. state-based conditions and intervals) from their original
formulations of \cite{dwyer98a} to remove assumptions of interleaving
and make them consistent with the left-closed, right-open intervals.
For brevity, this is the only pattern that we detail in this paper. We
are currently working on integrating our work into the pattern system
of Dwyer and his colleagues.  Meanwhile, the complete set of formulas
we developed is available at
{\verb$http://www.cs.toronto.edu/~chechik/edges.html$}.

\begin{figure*}[t]
  \begin{center}
    \fbox{
      \begin{minipage}{\linewidth}
        \(
        \begin{array}{rl}
          \multicolumn{2}{c}{$A. \(P\) Exists Globally$} \\
          0.&\e P \\
          1.&\e P \\
          2.&\e\eu P \\
          3.&\e\eu P
        \end{array}
        \)
        \hfill
        \(
        \begin{array}{rl}
          \multicolumn{2}{c}{$B. \(P\) Exists Before \(R\)$} \\
          0.&\e R \Limp \lnot(\lnot P \U R) \\
          1.&\e\eu R \Limp (\lnot\eu R \U P) \\
          2.&\e R \Limp \lnot(\lnot\eu P \U R) \\
          3.&\e\eu R \Limp \lnot(\lnot\eu P \U \eu R)
        \end{array}
        \)
        \hfill
        \(
        \begin{array}{rl}
          \multicolumn{2}{c}{$C. \(P\) Exists After \(Q\)$} \\
          0.&\e Q \Limp \e(Q \land \e P) \\
          1.&\e\eu Q \Limp \e(\eu Q \land \o\e P) \\
          2.&\e Q \Limp \e(Q \land \e\eu P) \\
          3.&\e\eu Q \Limp \e(\eu Q \land \e\eu P)
        \end{array}
        \) 
        \vspace{3mm}\par\noindent
        \(
        \begin{array}{rl}
          \multicolumn{2}{c}{$D. \(P\) Exists Between \(Q\) and \(R\)$} \\
          0.&\a(Q \land \e R \Limp \lnot(\lnot P \U R) \land \lnot R) \\
          1.&\a(\eu Q\land \e\eu R \Limp \o(\lnot\eu R\U P) \land\lnot\eu R) \\
          2.&\a(Q \land\e R \Limp \lnot(\lnot\eu P \U R) \land \lnot R) \\
          3.&\a(\eu Q\land \e\eu R\Limp\lnot(\lnot\eu P\U\eu R)\land\lnot\eu R)
        \end{array}
        \)
        \hfill
        \(
        \begin{array}{rl}
          \multicolumn{2}{c}{$E. \(P\) Exists After \(Q\) Until \(R\)$} \\
          0.&\a(Q \Limp \lif{\e R}{\lnot(\lnot P \U R)\land \lnot R}{\e P}) \\
          1.&\a(\eu Q \Limp \o(\lnot\eu R \U P) \land \lnot\eu R)\\
          2.&\a(Q\Limp\lif{\e R}{\lnot(\lnot\eu P\U R)\land\lnot R}{\e\eu P} \\
          3.&\a(\eu Q \Limp  \lif{\e\eu R}
                      {\lnot(\lnot\eu P \U \eu R)\land \lnot\eu R}{\e\eu P})\\
        \end{array}
        \)
      \end{minipage}
      }
      \caption{Formulations of the \textbf{Existence} Pattern}
    \label{fig:patE}
  \end{center}
\end{figure*}

\subsection{Closure Under Stuttering and Patterns}

Our primary concern associated with the edge-based extension of the
pattern system is to analyze the newly-created formulas for the
closure under stuttering property.  Our interest in this problem is
twofold: first, we know that edge-based formulas can have practical
relevance only if they are closed under stuttering, and second, these
formulas provide a good test-bed for the closure under stuttering
theorems we have developed.

Let us consider an example: a robot must pick up a metal blank from a
feed belt, weigh it, and deposit it in a press. The specification of
the robot says:
\begin{quote}
\em The robot must weigh the blank after pickup from the feedbelt, 
    but before the deposit in the press.
\end{quote}
The robot is equipped with a magnet and a scale at the end of its
arm. The status of the magnet is reported through a state variable
$\nm{mgn}$ which is true when the magnet is on, while the scale
reports a successful weighing when the state variable $\nm{scl}$ is
true.

Clearly, this fits the ``Existence'' pattern: a state base condition
(weighing) must happen between two events (pickup and deposit). In
Figure~\ref{fig:patE}, we can find the desired formula as D.1.  Using
$\eu\nm{mgn}$ and $\ed\nm{mgn}$ to model the pickup and
the deposit event, respectively, and plugging the events into the
template yields the formula: 
\[
  \a(\eu\nm{mgn}\land \e\ed\nm{mgn} \Limp
  \o(\lnot\ed\nm{mgn}\U \nm{scl}) \land \lnot\ed\nm{mgn})
\]
In order to prove that this property is closed under stuttering,
we need to show that if components of the template, $Q$, $R$ and $P$
are closed under stuttering, then so is the template. $Q$, $R$ and $P$ are
just variables, trivially closed under stuttering, and the analysis
of the template appears below: 
\begin{proof}
     & \cus{\a(\eu Q\land \e\eu R \Limp \o(\lnot\eu R\U P) \land\lnot\eu R)} 
       \lhint{by the laws of logic and LTL}
\Leq & \cus{\a(\eu Q\land \e\eu R \Limp \o(\lnot\eu R\U P)) \land \\
     & \phantom{\cusleft} \a(\eu Q\land \e\eu R \Limp \lnot\eu R)} 
       \lhint{by~\ref{eq:cusAnd}}
\Lpmi& \cus{\a(\eu Q\land \e\eu R \Limp \o(\lnot\eu R\U P))} \land \\
     & \cus{\a(\eu Q\land \e\eu R \Limp \lnot\eu R)}
       \lhint{by Property~\ref{eq:cusAe}, we get:}
\Lpmi& \cus{Q} \land \cus{\e\eu R} \land \cus{\lnot\eu R\U P} \land \\
     & \cus{\a(\eu Q\land \e\eu R \Limp \lnot\eu R)}
       \lhint{by Property~\ref{eq:cusEe}, this simplifies to:}
\Lpmi& \cus{Q} \land \cus{R} \land \cus{\lnot\eu R\U P} \land \\
     & \cus{\a(\eu Q\land \e\eu R \Limp \lnot\eu R)}
       \lhint{by Property~\ref{eq:cusUe}, we get:}
\Lpmi& \cus{Q} \land \cus{R} \land \cus{R} \land \cus{P} \land \\
     & \cus{\a(\eu Q\land \e\eu R \Limp \lnot\eu R)}
       \lhint{by the rules of logic we get:}
\Leq & \cus{Q} \land \cus{R} \land \cus{P} \land \\
     & \cus{\lnot\e(\eu Q\land \e\eu R \land \eu R)}
       \lhint{by~\ref{eq:cusNot} and Property~\ref{eq:cusEe}:}
\Lpmi& \cus{Q} \land \cus{R} \land \cus{P} \land \\
     & \cus{Q} \land \cus{R} \land \cus{\e\eu R}
       \lhint{by Property~\ref{eq:cusEe} again:}
\Lpmi& \cus{Q} \land \cus{R} \land \cus{P} \land \\
     & \cus{Q} \land \cus{R} \land \cus{R}
\end{proof}
Thus, we have proven that
\begin{multline}
\cus{P}\land\cus{Q}\land\cus{R} \Limp \\
\cus{\a(\eu Q\land \e\eu R \Limp \o(\lnot\eu R\U P) \land\lnot\eu R)} \nonumber
\end{multline}
which is exactly the desired property.

Note that, although the property is fairly complicated, the proof is
not long, is completely syntactic, and each step in it is easy.
Similar proofs were found for all of the new edge-based
formulas~\cite{paun99b}.  Such proofs can potentially be performed by a
theorem-prover like PVS~\cite{shankar93} with little guidance from the
user.  We are currently investigating the feasibility of doing so.

\section{Discussion and Related Work}
\label{discussion}

Before writing this paper, we searched through numerous research
publications and web sites, looking for good examples of LTL formulas
containing the ``next'' state, but surprisingly found just a few.  We
also looked at over 500 temporal formulas collected by Dwyer and his
colleagues~\cite{dwyer-web,dwyer99} and found virtually no explicit
usage of events or the ``next'' operator.  This led us to conclude
that the community is almost religiously avoiding the ``next'' state
operator, replacing it with a variety of surrogates, most of
which are neither elegant nor expressive.  

For example, to simulate an up edge, it is customary to create an
extra variable, $\hat{a}$, that signals a change in $a$ by being
temporarily true when $a$ is changed.  This can be modeled by
a concurrent assignment to $a$ and $\hat{a}$:
\[ 
  \mathbf{atomic}\{\hat{a} \leftarrow 1; a \leftarrow 1;\}\ 
  \hat{a} \leftarrow 0;
\]
Note that $\hat{a}$ is weaker than $\eu a$ since in systems containing
more than one process, $\hat{a}$ is true for a longer time than $\eu
a$.  Since $\eu a \limp \hat{a}$, replacing $\eu a$ by $\hat{a}$ in
LTL properties can lead to hard-to-interpret verification results.
For example, a property
\[
\a(\hat{a} \limp \e p)
\]
is a \emph{conservative} approximation of 
\[
\a(\eu a \limp \o\e p)
\]
That is, if the former property holds, so does the later, but the
converse is not true. In many such cases the approximation is too
strong, forcing the user to modify a possibly correct model just to be
able to verify the property. Such approximations can be so strong that
they are always false in most models. For example, performing the
substitution in the formula:
\[
\a(\eu a \limp \o(\lnot \eu a \U \eu b))
\]
will most likely result in a formula that is always false because
$\hat{a}$ can be true for several consecutive states.  A property
\[
\e(\hat{a} \land \hat{b})
\]
is an \emph{optimistic} approximation of
\[
\e(\eu a \land \eu b)
\]
That is, if the former property does not hold, neither does the later,
but the converse is not true. This means that the approximation cannot
be used for checking the validity of the model. Combining conservative
and optimistic approximations can void the resulting formula of any
meaning.

The users of the model-checker SPIN~\cite{holzmann97} probably
constitute the largest LTL user-group\footnote{SPIN has over
4000 installations world-wide.}.  However, judging from the four years
of proceedings of the SPIN workshop, available at \cite{spin-web},
they seldom if ever use properties involving events because of the
common misconception that no formulas containing ``next'' are closed
under stuttering, expressed, e.g., by Kamel and Leue in
\cite{kamel98}. This is, in our opinion, a serious problem because we
believe that edges are required to express most nontrivial
properties. For example, during our work~\cite{paun98} on the
Production Cell~\cite{lewerentz95}, edges were required in 10 out of
14 properties that we formalized, and in many of these, simulating
edges by introducing extra variables was not possible.

A restricted use of ``next'', similar to ours, is also advocated by Lamport in
his Temporal Language of Actions (TLA)~\cite{lamport94}, where
``next'' is replaced by ``primed variables'', e.g., $a'$ indicates the
value of $a$ in the next state. However, this is not sufficient to
guarantee closure under stuttering and an additional restriction is
placed on the TLA formulas. This restriction is similar in form to the
one imposed by Theorem~\ref{thm:main}.

\section{Summary and Conclusion}
\label{conclusion}

The ``next'' operator in linear-time temporal logics is required for
reasoning about events.  However, it is seldom if ever used in
practice because of a false belief that it does not allow construction
of formulas that are closed under stuttering.  Instead, people
introduce extra variables to simulate events.  These variables clutter
the model and make it harder to analyze.  Moreover, results of the
verification with respect to these properties often cannot be
interpreted correctly without complete understanding of the modeling
language and logic, leading to errors among novice and even expert
users.

In this paper, we have introduced the notion of edges in the context
of LTL, a concept that allows us to easily express temporal properties
involving events.  We have also provided a number of theorems that
enable syntax-based analysis of a large class of formulas for closure
under stuttering.  These theorems can be easily added to a theorem
prover for mechanized checking.  In addition, we extended the patterns
identified in~\cite{dwyer98a} with event-based formulations, and
proved that the resulting formulas are closed under stuttering using
the theorems presented in this paper.  Unfortunately, unlike the
``next''-free LTL, our language of edges is not closed, i.e., it is
possible to use this language to state a property which is not closed under
stuttering.  However, we feel that it can express and enable analysis
of the majority of formulas that are encountered in practice.  For
example, we were easily able to check the formalization of properties
of the Production Cell System.

We hope that the work presented in this paper will contribute to
increasing the usability of formal methods, at least in the
linear-temporal logic domain.

\vspace{6mm}

\noindent {\large\bf Acknowledgments}

\vspace{4mm}

We would like to thank Rick Hehner, Connie Heitmeyer, and the
anonymous referees for their helpful comments on the earlier drafts of
this paper.  The research was supported in part by the Natural Science
and Engineering Research Council of Canada.

\end{document}